\documentclass[journal,twocolumn]{IEEEtran}
\usepackage{xcolor}
\usepackage{mathtools}
\usepackage{epsfig,makeidx,color,subfigure}
\usepackage{amsmath,amssymb,bbm,enumitem}
\usepackage{amsthm}
\usepackage{cite,graphicx,lipsum}
\usepackage{algorithm}
\usepackage{algpseudocode}
\usepackage[switch,pagewise]{lineno}
\usepackage{hyperref}
\hypersetup{
        colorlinks = true,
        citecolor=red,
}

\def\bpsi{\mbox{\boldmath $\psi$}}

\def\cG{{\cal G}}

\def\rT{{\rm T}}

\def\uR{{\mathbb R}}

\def\uE{{\mathbb E}}

\DeclareMathOperator*{\argmin}{\arg\!\min}

\newtheorem{mytheorem}{\bf Theorem} 
\newtheorem{mylemma}{\bf Lemma} 

\def\be{ \begin{equation} }
\def\ee{ \end{equation} }
\def\bea{ \begin{eqnarray} }
\def\eea{ \end{eqnarray} }

\def\bx{{\bf x}}
\def\by{{\bf y}}
\def\bc{{\bf c}}

\def\bb{{\bf b}}

\def\bg{{\bf g}}

\def\ba{{\bf a}}
\def\br{{\bf r}}

\def\bm{{\bf m}}
\def\bn{{\bf n}}

\def\bv{{\bf v}}

\def\bA{{\bf A}}
\def\bB{{\bf B}}
\def\bC{{\bf C}}

\def\bF{{\bf F}}

\def\bI{{\bf I}}

\def\bK{{\bf K}}

\def\bW{{\bf W}}

\def\bX{{\bf X}}

\def\bone{{\bf 1}}

\def\b0{{\bf 0}}

\def\bPsi{{\bf \Psi}}

\def\bSigma{{\bf \Sigma}}

\def\cC{{\cal C}}

\def\cN{{\cal N}}

\begin{document}

\title{Iterative Sparse Identification of Nonlinear Dynamics}

\author{Jinho Choi \thanks{The author is with
the School of Information Technology,
Deakin University, Geelong, VIC 3220, Australia
(e-mail: jinho.choi@deakin.edu.au).}\\
}

\maketitle
\begin{abstract}
In order to extract governing equations from time-series data, various approaches are proposed. Among those, sparse identification of nonlinear dynamics (SINDy) stands out as a successful method capable of modeling governing equations with a minimal number of terms, utilizing the principles of compressive sensing. This feature, which relies on a small number of terms, is crucial for interpretability. The effectiveness of SINDy hinges on the choice of candidate functions within its dictionary to extract governing equations of dynamical systems. A larger dictionary allows for more terms, enhancing the quality of approximations. However, the computational complexity scales with dictionary size, rendering SINDy less suitable for high-dimensional datasets, even though it has been successfully applied to low-dimensional datasets. To address this challenge, we introduce iterative SINDy in this paper, where the dictionary undergoes expansion and compression through iterations. We also conduct an analysis of the convergence properties of iterative SINDy. Simulation results validate that iterative SINDy can achieve nearly identical performance to SINDy, while significantly reducing computational complexity. Notably, iterative SINDy demonstrates effectiveness with high-dimensional time-series data without incurring the prohibitively high computational cost associated with SINDy.
\end{abstract}


\section{Introduction}

When presented with a dataset as time-series, one of the primary objectives is to gain a comprehensive understanding of the underlying phenomena it represents. It is often useful to discover dynamics governing the observed behavior. In this pursuit, the identification or discovery of nonlinear dynamics plays a pivotal role \cite{Brunton22} \cite{Brunton_Kutz_2022}. By identifying nonlinear patterns, relationships, and dependencies within the data, we can elucidate complex behaviors that might otherwise remain hidden, which has been studied in fluid mechanics \cite{Schmid10} \cite{Tu14} \cite{Williams15}, biophysical processes \cite{Boninsegna18},
and forecasting sequential data \cite{Azencot20a} \cite{Lange21}.
This process enables us to construct meaningful models, make accurate predictions, and extract valuable insights from the dataset.

There are data-driven techniques that can discover modes of dynamics, such as dynamic mode decomposition (DMD) \cite{Schmid10} \cite{Tu14} and extended DMD (EDMD) \cite{Williams15} under certain assumptions. In particular, in DMD, it is assumed that the measurements of time-series have the following relationship:
\be 
\bx (t+1) \approx \bA \bx(t) \in \uR^N,
    \label{EQ:DMD}
\ee 
where $\bx(t)$ represents a sample of dimension $N$ obtained at time $t$ and $\bA \in \uR^{N \times N}$ is an unknown square matrix. 
Here, throughout the paper, matrices and vectors are denoted by upper- and lower-case boldface letters, respectively.
From a given dataset, $\bA$ is estimated and its eigenvectors are called modes. 

In EDMD, it is assumed that the next measurement, say $\bx(t+1)$, can be approximated using a nonlinear function of the current measurement, $\bx(t)$. That is,
\be 
\bx(t+1) = \bF (\bx(t)) \in \uR^N,
    \label{EQ4:xFx_k}
\ee 
where $\bF: \ \uR^N \to \uR^N$ represents a nonlinear function. This assumption is particularly natural when the measurements represent the state vector of a nonlinear dynamical system. Within the context of nonlinear dynamical systems, the nonlinear function $\bF$ is referred to as the flow map. Then, based on the Koopman representation  \cite{Koopman31} \cite{Budii12} \cite{Brunton_PLOS}, a vector of observables, denoted by $g_1 (\bx), \ldots, g_M (\bx)$, is assumed to have the following relationship:
\be 
\bg (t+1) \approx \bK \bg(t) \in \uR^M,
    \label{EQ:gKg}
\ee 
where $\bg(t) = [g_1 (\bx(t)) \ \ldots \ g_M (\bx(t))]^\rT$ and $\bK \in \uR ^{M \times M}$ is called the Koopman matrix. 
Here, the superscript $\rT$ denotes the transpose.
In general, the relationship in \eqref{EQ:gKg} is an approximation with a finite value of $M$ \cite{Brunton_PLOS}. The choice of observable functions strongly influences the approximation error and often requires to have prior knowledge of measurements or time-series datasets.

In general, when applying data-driven techniques like DMD or EDMD, the primary objective is typically to reveal the governing linear or nonlinear dynamics responsible for generating these measurements, respectively.

In \cite{Brunton16}, a different data-driven technique, called sparse identification of nonlinear dynamics (SINDy), is proposed for the direct identification of the flow map in \eqref{EQ4:xFx_k}. 
This flow map is approximated through a linear combination of a few candidate vectors chosen from an extensive dictionary by using a compressive sensing technique \cite{Eldar12} \cite{Donoho06} \cite{Candes06}. 
This dictionary comprises observable functions, which are memoryless nonlinear functions. Recognizing that datasets may correspond to any nonlinear dynamics, the inclusion of a substantial number of candidate functions in the dictionary emerges as a critical step for effectively capturing the intricate underlying nonlinear dynamics. Nevertheless, the computational complexity increases as the dictionary size grows, which constrains the applicability of SINDy to high-dimensional time-series datasets, as the dictionary size can quickly increase with the input data dimension \cite{Baddoo22}.

In this paper, we propose an iterative approach to SINDy that expands and compresses the dictionary through iterations, starting with an initially small dictionary. The iterative SINDy approach does not rely on a large dictionary for approximations but adjusts the dictionary's size by expanding and compressing it through iterations for a given time-series dataset. As a result, its computational complexity is relatively low compared to that of conventional SINDy, making it suitable for handling high-dimensional datasets. We also conduct an analysis of iterative SINDy to understand its convergence properties and demonstrate that it converges when the sparsity constraint weight is sufficiently large. Our numerical results reveal that the performance in terms of approximation is comparable to that of SINDy, while the computational complexity can be significantly lower, particularly when the input dimension is not low.

The remainder of the paper is structured as follows: In Section~\ref{S:Background}, we provide background information, including an overview of SINDy. Section~\ref{S:ISINDy} introduces our proposed iterative approach to SINDy, in which the dictionary is adaptively formed by iteratively expanding and compressing it using monomials of input variables. Convergence properties are derived in Section~\ref{S:Analysis}.
We present our numerical results in Section~\ref{S:Sim}. Finally, the paper is concluded with remarks in Section~\ref{S:Conc}.

\subsubsection*{Notation}
$||\cdot||_p$ represents the $p$-norm and $||\cdot||_{\rm F}$ stands for the Frobenius norm.
The identity matrix is denoted by $\bI$. $\bone_{n}$ represents
the $n \times 1$ column vector of all ones.
$\uE[\cdot]$ and ${\rm Var}(\cdot)$
denote the statistical expectation and variance, respectively.
$\cN(\bm, \bSigma)$ represents the Gaussian distribution with mean $\bm$ and covariance matrix $\bSigma$.

\section{Background}    \label{S:Background}

\subsection{Compressive Sensing}

In compressive sensing, a representation of a given signal is to find as a linear combination of a sparse subset of candidate functions derived from a comprehensive dictionary containing a multitude of potential candidates \cite{Candes06} \cite{Donoho06}. The fundamental premise of compressive sensing hinges on the idea that many real-world signals exhibit inherent sparsity, meaning that only a small number of candidate functions from the dictionary are necessary to faithfully capture their essential characteristics. 

Consider a signal represented as a vector $\by$ of length $D$. We define a dictionary $\cC$ consisting of $J$ candidate vectors, denoted as $\cC = \{\bc_1, \ldots, \bc_J\}$, with each $\bc_j$ being a vector of length $D$. For simplicity, we can also express the dictionary as $\bC = [\bc_1\ \ldots \ \bc_J]$, employing a slight abuse of notation. In general, $J \gg D$, and the goal is to find a subset of candidate vectors that can approximate $\by$ as follows:
\be 
\by \approx \sum_{q = 1}^Q b_{j(q)} \bc_{j(q)} ,
    \label{EQ:sest}
\ee
where $j(q) \in \{1, \ldots, J\}$ is the index of the $q$th selected candidate vector, and $b_{j(q)}$ is its corresponding coefficient. The parameter $Q\ (\ll J)$ represents the sparsity, i.e., the number of selected candidate vectors.

Let $\bb$ be a weight vector of length $J$, and denote by $|| \bx ||_0$ the $\ell_0$-norm, which represents the number of non-zero elements in $\bx$. Then, the problem of finding a sparse solution with sparsity $Q$ to approximate a target signal $\by$ as described in \eqref{EQ:sest}, can be formulated as follows:
\be 
\hat \bb = \argmin_{\bb: \ ||\bb||_0 = Q} ||\by - \bC \bb||_2^2 .
\ee 
Solving this problem through an exhaustive search becomes computationally prohibitive as the size of dictionary, $J$, increases. Therefore, various low-complexity approaches have been proposed to find a sparse solution by selecting a subset of candidate vectors from the dictionary in compressive sensing \cite{Eldar12} \cite{Foucart13}. Among these methods, the Least Absolute Shrinkage and Selection Operator (Lasso) \cite{Tibshirani96} is one of the widely used and effective techniques. The Lasso method operates by minimizing a weighted sum of the least-squares error and a regularization term, typically the $\ell_1$-norm of the coefficient vector, represented as follows:
\be
\hat \bb =\argmin_{\bc}  ||\by - \bC \bb||_2^2 + \beta ||\bb||_1,
    \label{EQ:21}
\ee
where $\beta > 0$ is the Lagrange multiplier or weight for the regularization term.  

In \eqref{EQ:21}, the $\ell_1$-norm is utilized for regularization to promote sparsity in the solution vector $\bb$. This choice does not affect the convexity of the problem defined in \eqref{EQ:21}, as both the least-squares error term and the $\ell_1$-norm regularization term are convex functions, and the minimization problem is over a convex domain \cite{Foucart13}. 
There are also other techniques based on sparse least squares with Bayesian frameworks as studied\footnote{An anonymous reviewer provided the related references.} in \cite{Schaeffer2017} \cite{Tripura2023} \cite{Nayek2021} \cite{Chen2021}. In particular, the approach in \cite{Tripura2023} can be considered to extend the approach to stochastic dynamical systems.

\subsection{Sparse Identification of Nonlinear Dynamics (SINDy)}

SINDy is to find a solution to approximately express the next state in terms of the current state based on the assumption that the approximate is a linear combination of a few candidate functions in  a large dictionary \cite{Brunton16}.   

Let $\psi_k (\bx)$ be the $k$th candidate function of the dictionary, which is a function of state vector $\bx$.
Denote by $\bpsi(\bx) = [\psi_1 (\bx) \ \ldots \ \psi_K (\bx)]^\rT$ the vector of the candidate functions, where $K$ is the size of the dictionary. Suppose that the right-hand side (RHS) term in \eqref{EQ4:xFx_k} is approximated as
\begin{align} 
[\bF (\bx )]_n \approx \bb_n^\rT \bpsi (\bx) = \bpsi (\bx)^\rT \bb_n, \ n = 1,\ldots, N,
    \label{EQ4:Fxn}
\end{align}
where $\bb_n = [b_{n,1} \ \ldots \ b_{n, K}]^\rT$ is a sparse vector. In other words, each element of the next state vector is approximated by a linear combination of a few elements of the dictionary as follows:
\begin{align} 
x_n (t+1) & = [\bF (\bx (t))]_n \cr 
& \approx \bpsi (\bx(t))^\rT \bb_n. 
\end{align}
This representation ensures that only a concise set of candidate functions, represented by the non-zero elements of $\bb_n$, actively participate in predicting the evolution of $x_n(t+1).$ Such a sparse representation allows for a more parsimonious and interpretable model of the system's dynamics \cite{Brunton16}. Consequently, the primary task becomes finding the optimal sparse structure for $\bb_n$, where the nonzero coefficients identify the most relevant terms from the dictionary of candidate functions.

We assume that a set of measurements, $\{\bx(0), \ldots, \bx (T)\}$, is available as a multivariate time-series dataset. 
In addition, let
\begin{align}
\bar \bX & = [ \bx(1) \ \ldots \bx (T)] \in \uR^{N \times T} \cr 
\bPsi & = [\bpsi(\bx(0)) \ \ldots \bpsi (\bx(T-1))] \in \uR^{K \times T}.
\end{align}
Substituting \eqref{EQ4:Fxn} into \eqref{EQ4:xFx_k}, we have
\be 
\bx (t +1) \approx \bB^\rT \bpsi (\bx (t)), \ t = 0, \ldots, T-1,
\ee 
or 
\be 
\bar \bX \approx \bB^\rT \bPsi ,
    \label{EQ:ASINDy}
\ee 
where $\bB = [\bb_1 \ \ldots \ \bb_N] \in \uR^{K \times N}$. 
Note that if $\psi_k (\bx) = x_k$, $k = 1, \ldots, K$, where $K = N$, \eqref{EQ:ASINDy} becomes \eqref{EQ:DMD} with $\bB^\rT = \bA$
(i.e., SINDy can be seen as a generalization of DMD). 
Now, $\bB$ can be determined to minimize $||\bar \bX - \bB^\rT \bPsi||_{\rm F}^2$ while imposing sparsity constraints. 
Note that
\be 
||\bar \bX - \bB^\rT \bPsi||_{\rm F}^2 = \sum_{n=1}^N ||\bar \bx_n - \bPsi^\rT \bb_n||_2^2,
\ee 
where 
\be 
\bar \bx_n = [x_n (t_1+1) \ \ldots \ x_n (t_T +1)]^\rT \in \uR^T.
\ee 
Thus, minimizing $||\bar \bX - \bB^\rT \bPsi||_{\rm F}^2$ can be decomposed into $M$ subproblems, each of which can be expressed as
\be
\min_{\bb_n} ||\bar \bx_n - \bPsi^\rT \bb_n||^2 \ \text{subject to constraints on sparsity}.
    \label{EQ4:l2_spar}
\ee

The solution to the problem in \eqref{EQ4:l2_spar} depends on the specifics of the sparsity constraints. In practice, these constraints can take various forms, such as $\ell_1$-regularization (i.e., Lasso), which encourages a sparse solution by adding an absolute value penalty term to the objective function. That is, with a penalty term to encourage a sparse solution, \eqref{EQ4:l2_spar} can be rewritten as
\be
\min_{\bb_n} ||\bar \bx_n - \bPsi^\rT \bb_n||^2_2 + \beta ||\bb_n||_1.
    \label{EQ4:lasso}
\ee
The penalty term, expressed as $||\bb_n||_1$, sums up the absolute values of the coefficients residing within the parameter vector $\bb_n$. During the optimization process driven by this penalty term, the algorithm actively seeks to minimize it. As a result, it tends to drive many coefficients to precisely zero. This behavior results in a sparse solution, where only a subset of coefficients remains non-zero, effectively selecting a limited set of candidate functions from the dictionary. 
The problem in \eqref{EQ4:lasso} can be solved using any convex solver as the objective function is convex \cite{Boyd}. The resulting approach is called the SINDy approach.

\section{Iterative SINDy}   \label{S:ISINDy}

The utilization of a large dictionary in EDMD and SINDy can  improve modeling accuracy and provide increased flexibility when capturing the underlying dynamics of a system. This is particularly beneficial when dealing with complex or high-dimensional data, as it allows for a richer representation of the system's behavior. However, it is noteworthy that the use of a large dictionary simultaneously introduces computational challenges due to its size and the associated computational complexity. In this section, we propose an iterative approach to perform SINDy where the size of dictionary in each iteration can be limited.

\subsection{Dictionary Expansion and Compression} \label{SS:Example}

In this subsection, we demonstrate, with examples, how to expand and compress the dictionary of candidate functions, a crucial aspect in iterative approaches for SINDy.

Consider the logistic map as a simple nonlinear dynamical systems, which is given by
\begin{align}
x (t+1) & = r x(t) (1-x(t)) \cr 
& = F (x(t)),
\end{align}
where $F (x) = r x (1-x) = -rx^2 + rx$.
Let 
$$
\bar \bx = [x(1) \ \ldots \ x (T)]^\rT \in \uR^{1 \times T}.
$$
With multiple monomials, i.e., $\psi_k (x) = x^{k-1}$, $k = 1, \ldots, K$, we can form $\bpsi (x)$. Let $K = 4$. Then, we can show that
\begin{align}
 \bar \bx & = \bb^\rT [\bpsi (x(0)) \ \ldots \ \bpsi (x(T-1))] \cr 
 & = [0\ r\ -r \ 0] \left[ 
 \begin{array}{cccc}
 1 & 1 & \cdots & 1 \cr 
 x(0) & x(1) & \cdots & x (T-1) \cr 
 x(0)^2 & x(1)^2 & \cdots & x (T-1)^2 \cr
 x(0)^3 & x(1)^3 & \cdots & x (T-1)^3 \cr
 \end{array}
 \right].   
    \label{EQ:bx4}
\end{align}
While the correct candidate functions, which are $\psi_2 (x) = x$ and $\psi_3 (x) = x^2$, can be chosen, it may have a high computational complexity with $K = 4$. To lower computational complexity, we can first consider the approximation with $K = 2$, i.e.,
$\psi_1 (x) = 1$ and $\psi_2 (x) = x$. Then, the following approximation is considered:
$$
\bar \bx \approx [b_1 \ b_2] \left[ 
 \begin{array}{cccc}
 1 & 1 & \cdots & 1 \cr 
 x(0) & x(1) & \cdots & x (T-1) \cr 
 \end{array}
 \right].
$$
In this case, neither $b_1$ nor $b_2$ is close to 0, which allows us to consider more candidate functions by expanding the dictionary. To achieve this, we introduce the operation of elementwise multiplication, denoted by $\boxtimes$. For instance, $\{a,b\} \boxtimes \{c,d\} = \{ac, ad, bc, bd\}$. It can be shown that $\{\psi_1(x), \psi_2(x)\} \boxtimes \{\psi_1(x), \psi_2(x)\} = \{\psi_1(x), \ldots, \psi_4(x)\}$. With this expanded candidate set, we can obtain \eqref{EQ:bx4}. Since the sparsity of $\bb$ is 2, only $\psi_2(x) = x$ and $\psi_3(x) = x^2$ will remain for the next iteration. The dictionary, i.e., the set of candidate functions, can be further expanded as follows:
\begin{align*}
\{\psi_2 (x), \psi_3(x)\} \boxtimes \{\psi_1 (x), \psi_2(x)\} 
& =  \{\psi_2 (x), \psi_3(x), \psi_4 (x)\} \cr 
& = \{x, x^2, x^3\}.    
\end{align*}
With this set of candidate functions, $\bx(t+1)$ can be approximated. Consequently, the coefficient vector becomes $\bb = [b_1 \ b_2 \ b_3]^\rT = [r \ -r \ 0]^\rT$, meaning that $\psi_2(x)$ and $\psi_3(x)$ 
are once again identified as the survivors, and further iterations will not alter this selection.

In the above example, we observe that the coefficient vectors are not sparse when all the candidate functions are required to approximate $\bx(t+1)$. Consequently, in such cases, it becomes necessary to expand the set of candidate functions. On the other hand, if the dictionary is sufficiently large and some elements of the coefficient vectors can be disregarded, the set can be compressed. By iteratively expanding and compressing the set of candidate functions, we could ultimately identify a subset of candidate functions that effectively approximates $\bx(t+1)$.

\begin{figure}[!t]
\begin{center} \hspace{-1.5em} 
\includegraphics[width=9cm]{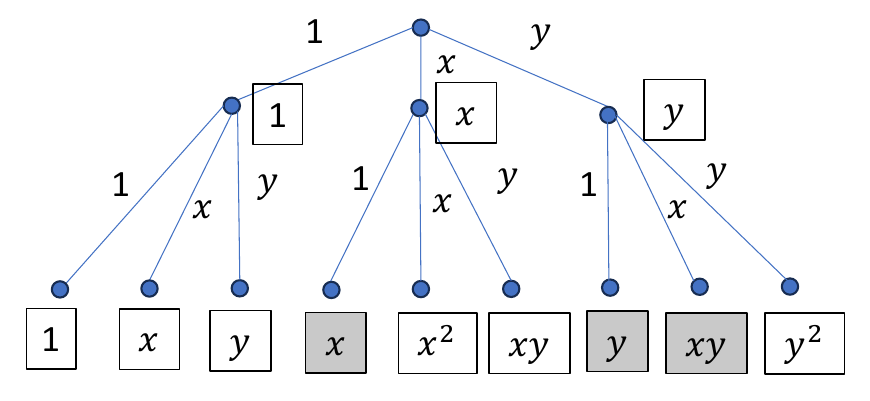}
\end{center}
\caption{Dictionary expansion when $N = 2$  (the shaded boxes represent the redundant monomials due to duplications, which can be removed).}
    \label{Fig:1xy}
\end{figure}

We can consider another example when $N = 2$ with a dictionary of monominals in Fig.~\ref{Fig:1xy}. 
By repeatedly using the operation
of elementwise multiplication, the dictionary can be expanded with more monomials of higher degrees. It can be readily shown that this expansion can be carried out with any $N \ge 1$. The expanded dictionary can be compressed to keep a few survivor monomials by using compressive sensing as in \eqref{EQ4:l2_spar}.

\subsection{The Proposed Approach}


If the flow map $\bF$ in \eqref{EQ4:xFx_k} is Lipschitz continuous and can be approximated effectively by a polynomial function, we can employ a dictionary of monomials for SINDy. As demonstrated in Subsection~\ref{SS:Example}, the dictionary can be expanded and compressed through iterations. In this subsection, for any $N \ge 1$, we introduce an approach for iterative SINDy with monomials to approximate the polynomial function. This approach involves both expanding and compressing the dictionary.

We begin by defining $L$ initial candidate functions as follows:
\be 
\cG_1 = \{g_1 (\bx), \ldots, g_L (\bx)\},
\ee 
where $L$ is chosen such that $L \ge N$. In a specific instance, we can set $\cG_1 = \{1, x_1, \ldots, x_N\}$ with $L = N+1$ as the unity set. It is important to note that each element of $\cG_1$ is either 1 (representing a monomial of degree 0) or a linear term (representing a monomial of degree 1). 
Using the unity set, the dictionary as the set of candidate functions can be expanded to include monomials of higher degrees as follows:
\be
\tilde \cG_{s} = \cG_{s-1} \boxtimes \cG_1, \ s \ge 2.
    \label{EQ4:GG}
\ee
In this expression, $\tilde \cG_s$ contains monomials of higher degree than those present in $\cG_{s-1}$, demonstrating the effectiveness of elementwise multiplication. In particular, it can be readily shown that $\cG_s$ can contain monomials of degrees up to $s$.
To illustrate, consider a specific case where $N = 2$ and $\cG_2 = \{x_1^2, x_2^2, x_1 x_2^5\}$. From \eqref{EQ4:GG}, we can show that
$$
\tilde \cG_3 = \{x_1^2, x_1^3, x_1^2 x_2, x_2^2, x_1 x_2^2, x_2^3, x_1 x_2^6, x_1^2 x_2^5, x_1 x_2^6 \},
$$

While it is possible to expand the dictionary through iterations with $\cG_s = \tilde \cG_s$, this approach can result in a growing dictionary size. To address this concern, we employ SINDy in each iteration to select a few surviving terms.  To this end, denote by $\bpsi_s (\bx)$  the vector of candidate functions in $\tilde \cG_s$ as $\bpsi_s (\bx)$. For example, if $\tilde \cG_2 = \{x_1^2, x_2^2, x_1 x_2^5\}$, then
$$ 
\bpsi_3 (\bx) = [x_1^2 \ x_2^2 \ x_1 x_2^5]^\rT.
$$
In iteration $s$, we solve the optimization problem:
\be
\min_{\bb_n} ||\bar \bx_n - \tilde \bPsi_s^\rT \bb_n||^2 \ \text{subject to constraints on sparsity},
    \label{EQ4:ss1}
\ee
where $\tilde \bPsi_s$ represents the matrix of candidate functions associated with $\tilde \cG_s$ or $\bpsi_s (\bx)$. The outcome is a set of candidate functions, referred to as the survivors, which form $\cG_s$. Clearly, $\cG_s$ is a subset of $\tilde \cG_s$. The corresponding matrix is denoted by $\bPsi_s$, which is also a submatrix of $\tilde \bPsi_s$.
This approach allows us to control the dictionary's size while capturing the most relevant terms from the expanded set of candidate functions for modeling.

Through iterations, we can identify a subset of monomial terms from $\bx(t)$ to approximate $x_n(t+1)$ using a linear combination of these terms. The process in \eqref{EQ4:GG} involves expanding the dictionary, while the step in \eqref{EQ4:ss1} is to compress the dictionary by selecting a subset of terms during each iteration. The resulting algorithm  can found in Algorithm~\ref{ALG:ISINDy}, where $S$ is a parameter that determines the maximum degree of monomials in the dictionary.


\begin{algorithm}
\caption{Iterative SINDy}
    \label{ALG:ISINDy}
\begin{algorithmic}[1]
\State Initialize $\cG_1 = \{1, x_1, \ldots, x_N\}$ \Comment{Initial candidates}
\For{$s = 2$ to $S$} \Comment{Iterate for desired number of steps}
    \State Expand candidates: $\tilde \cG_{s} = \cG_{s-1} \boxtimes \cG_1$ \Comment{Expanding}
    \State Initialize $\bpsi_s (\bx)$ with candidates in $\tilde \cG_s$
    \For{$n = 1$ to $N$} 
        \State Solve optimization problem:
        \State \hspace{1em} $\bb_n^\ast = \argmin_{\bb_n} ||\bar \bx_n - \tilde \bPsi_s^\rT \bb_n||_2^2$ subject to sparsity constraints
        \State Update $\cG_s$ with the surviving terms  \Comment{Compressing}
    \EndFor
\EndFor
\end{algorithmic}
\end{algorithm}

The complexity of iterative SINDy depends on the number of iterations and the complexity of finding $\bn^\ast$ in each iteration. If we consider the Lasso formulation in \eqref{EQ4:lasso}, its complexity is known to be $O(L_s^3 + L_s^2 T)$ when $T > L_s$ \cite{Efron2004}, where $L_s$ represents the number of candidate vectors in iteration $s$. Thus, if $L_s \le \bar L$, where $\bar L$ is the maximum size of the library, the total complexity is $O(S(\bar L^3 + \bar L^2 T))$. Note that in conventional SINDy, there is no iteration, and the Lasso problem is solved with a large dictionary. Thus, the complexity of SINDy is $O(K^3 + K^2 T)$, where $K > \bar L$. Therefore, with a sufficiently large $T \ (> K)$, the complexity of iterative SINDy can be lower than that of conventional SINDy by a factor of $\frac{K^2}{S \bar L^2}$. For example, if $K = 100$ and $\bar L = 10$, we have $\frac{K^2}{S \bar L^2} = \frac{100}{S}$, while the number of iterations, $S$, is usually less than $10$.

In iterative SINDy, the unity set, which includes monomials of degrees up to 1, and the operation of elementwise multiplication are of significant importance. These elements can play a critical role in the linear and nonlinear disambiguation within the context of SINDy \cite{Baddoo22}.

If the elements of $\bF (\bx (t))$ are polynomials of $x_n (t)$, we can determine a finite number of iterations, allowing us to decide the appropriate number in advance. However, if these elements are not polynomial or are unknown, determining the required number of iterations becomes challenging. Therefore, in practice, it is necessary to set a sufficiently large maximum number of iterations.
\section{Convergence Analysis}  \label{S:Analysis}


In this section, our objective is to analyze the iterative SINDy approach with the aim of elucidating its convergence properties. The iterative SINDy approach is notable for its dynamic manipulation of dictionary expansion and compression. Hence, it becomes crucial to determine the stopping criteria for this iterative process, shedding light on the algorithm's reliability and effectiveness in approximating nonlinear system dynamics. 

\subsection{With Sparsity Constraints}

For analysis, suppose that the Lasso algorithm \cite{Tibshirani96} is used to take into account sparsity constraints. With Lasso, from \eqref{EQ4:lasso}, the cost function to find a sparse solution for the $s$th iteration in iterative SINDy is given by
\begin{align} 
C_s (\bb) = ||\bar \bx_n - \tilde \bPsi_{s}^\rT \bb  ||_2^2 + \beta ||\bb||_1.
    \label{EQ:Cbb}
\end{align}
For convenience, let $\bb_n^{(s)}$ denote the solution of the $s$th iteration in \eqref{EQ:Cbb}. We can then establish the following result regarding the convergence of iterative SINDy.

\begin{mytheorem}
Suppose that $\bPsi_s = \bPsi_{s-1}$ for some $s$. Then, $\bPsi_{s+i} = \bPsi_s$ for all $i \ge 1$.
\end{mytheorem}
\begin{proof}
From \eqref{EQ4:GG}, we can demonstrate that
\begin{align}
\tilde \bPsi_{s}^\rT = [\bPsi_{s-1}^\rT \ \bW_s],
    \label{EQ:PPW}
\end{align}
where $\bW_s$ is the matrix consisting of the row vectors corresponding to the newly added candidate vectors after the elementwise multiplication.  Note that since $1$ is an element of the unity set, $\cG_1$, after the elementwise multiplication,  $\tilde \cG_s$ has all the monomials in $\cG_{s-1}$. Thus, $\bPsi_{s-1}^\rT$ has to be a submatrix of $\tilde \bPsi_{s}^\rT $ as in \eqref{EQ:PPW}.
Substituting \eqref{EQ:PPW} into \eqref{EQ:Cbb}, we obtain
\begin{align}
C_s(\bb) = ||\bar \bx_n - \bPsi_{s-1}^\rT \ba - \bW_s \br ||_2^2 + \beta ||\ba||_1 + \beta ||\br||_1 ,
\end{align}
where $\bb = [\ba^\rT \ \br^\rT]^\rT$. Consequently, it can be demonstrated that
\begin{align}
\min_\bb C_s (\bb)
= \min_{\ba, \br} ||\bar \bx_n - \bPsi_{s-1}^\rT \ba - \bW_s \br ||_2^2 + \beta ||\ba||_1 + \beta ||\br||_1.
    \label{EQ:mCs}
\end{align}
Therefore, if $\bPsi_s = \bPsi_{s-1}$, $\br$ must be a zero vector (because $\bPsi_s = \bPsi_{s-1}$ means that the row vectors of $\bW_s$ are not selected for the sparse solution of the minimization problem in \eqref{EQ:mCs}). 
With $\tilde \bPsi_{s+1}$, derived from \eqref{EQ:PPW}, it can also be shown that
$$
\tilde \bPsi_{s+1}^\rT = [\bPsi_{s}^\rT \ \bW_{s+1}] = [\bPsi_{s-1}^\rT \ \bW_{s+1}],
$$
indicating that the weight vector corresponding to $\bW_{s+1}$ becomes a zero vector again, and $\bPsi_{s+1} = \bPsi_s = \bPsi_{s-1}$. This completes the proof, since we can have the same result with $\bPsi_{s+i}$ for $i \ge 1$. 
\end{proof}

As a result, Algorithm~\ref{ALG:ISINDy} can be modified to include the condition $\bPsi_s = \bPsi_{s-1}$ to terminate the iteration
as a stopping criterion,  without requiring $S$ iterations.

It would also be interesting to determine when $\bPsi_s$ can be equal to $\bPsi_{s-1}$. To explore this, let $\ba = \bb_n^{(s-1)}$. Then, we have
\begin{align} 
C_s(\bb) & = ||\bar \bx_n - \bPsi_{s-1}^\rT \bb_n^{(s-1)} - \bW_s \br ||_2^2 \cr 
& \quad + \beta ||\bb_n^{(s-1)}||_1 + \beta ||\br||_1 .
    \label{EQ:Cs2}
\end{align}
Let $\bv = \bar \bx_n - \bPsi_{s-1}^\rT \bb_n^{(s-1)}$ and assume that $||\bv||_2^2$ is sufficiently small for some $s$.
Then, we can have the following result.
\begin{mylemma}
Consider the following problem:
\be 
\br^\ast (\bv) = \argmin_\br C (\br; \bv),
    \label{EQ:brv}
\ee 
where 
\be 
C (\br; \bv) = ||\bv - \bW_s \br ||_2^2 + \beta ||\br||_1.
\ee
Suppose that $||\bv||_2^2 \le \epsilon$, where $\epsilon > 0$.
There exists $\epsilon_0 > 0$ such that $\br^\ast = \b0$ for all $\epsilon \le \epsilon_0$. 
\end{mylemma}
\begin{proof}
First, consider the case where $\epsilon = 0$ or $\bv = \b0$. In this case, we have $\br^\ast = \b0$ because $\beta > 0$.
Now, assume that $\br^\ast \neq \mathbf{0}$, while $\epsilon > 0$. This implies that
\begin{align}
\min_\br  C (\br; \bv) 
& = ||\bv - \bW_s \br^\ast ||_2^2 + \beta ||\br^\ast||_1 \cr
& <  C (\b0; \bv)  = ||\bv||_2^2 \le \epsilon,
\end{align} 
because $\br = \b0$ cannot be the minimizer of $C (\br; \bv)$ due to the assumption. Since $||\bv - \bW_s \br^\ast ||_2^2 \ge 0$, we have 
\be 
\beta ||\br^\ast||_1 \le ||\bv - \bW_s \br^\ast ||_2^2 + \beta ||\br^\ast||_1  < \epsilon.
    \label{EQ:key}
\ee 
With $\beta > 0$, as $\epsilon$ remains within an upper bound $\epsilon_0$, when $\epsilon_0$  becomes smaller and approaches $0$,
the optimal solution $\br^\ast$ must be $\b0$, which contradicts the assumption that $\br^\ast \ne \b0$. Thus, the solution in \eqref{EQ:brv} has to be $\br^\ast = \b0$ for all $\epsilon \le \epsilon_0$, since $C (\br; \bv)$ is a continuous function of $\br$.
\end{proof}

In \eqref{EQ:Cs2}, recalling that $\bb = [\ba^\rT \ \br^\rT]^\rT$, $C_s (\bb)$ is to be minimized with respect to $\ba$ and $\br$. Assuming that $\ba = \bb_n^{(s-1)}$, when $||\bv||_2^2 = ||\bar \bx_n - \bPsi_{s-1}^\rT \bb_n^{(s-1)}||_2^2$ becomes sufficiently small, we can demonstrate that $\br^\ast$ becomes a zero vector. This implies that a small error term in $||\bv||^2_2 = ||\bar \bx_n - \bPsi_{s-1}^\rT \bb_n^{(s-1)}||_2^2$ serves as a sufficient condition for $\bPsi_s$ to be equal to $\bPsi_{s-1}$, because 
\be
\bb_n^{(s)} =\left[
\begin{array}{c} \bb_n^{(s-1)} \cr \br^\ast \end{array} \right] = \left[
\begin{array}{c} \bb_n^{(s-1)} \cr \b0 \end{array} \right].
\ee

\subsection{Without Sparsity Constraints}

In the previous subsection, we discussed scenarios in which $\beta$ is greater than zero. 
As $\beta$ increases, it places more emphasis on achieving a sparse solution. In extreme cases, we might set $\beta$ to a large value. When this occurs, as indicated by \eqref{EQ:key}, it is possible for $\br^\ast$ to become a zero vector, even when $||\bv||_2^2 = \epsilon$ is not small. Essentially, if $\beta \gg \epsilon$, $\br^\ast$ becomes a zero vector, implying that no further iterations are necessary. Furthermore, with $\br^\ast = \b0$, since $\beta$ is large, the optimal $\ba$ that minimizes the following cost function also becomes a zero vector:
$$ 
||\bar \bx_n - \bPsi_{s-1}^\rT \ba ||_2^2 + \beta ||\ba||_1.
$$ 
This means that for a large $\beta$, the optimal solution tends to be a zero-vector. To address this problem, it might be necessary to maintain a relatively small $\beta$. Consequently, it becomes intriguing to explore scenarios where $\beta \to 0$, which may result in non-sparse solutions.

With $\beta = 0$, let $K_s$ represent the number of rows in $\tilde \bPsi_s$. If $T > K_s$ and $\tilde \bPsi_s$ is full-rank, the optimal vector minimizing the cost function in \eqref{EQ:Cbb} is:
\be 
\bb_n^{(s)} = \left( \tilde \bPsi_s^\rT \right)^\dagger \bar \bx_n,
\ee 
where $\bA^\dagger$ denotes the pseudo-inverse of $\bA$ \cite{Golub1996matrix}. Since $\bb_n^{(s)}$ is not sparse, no compression of the dictionary in iterative SINDy will occur, meaning that the size of the dictionary, $K_s$, will grow through iterations. Thus, with a small $\beta$ or weak sparsity constraints, it would be difficult to find a compact model to approximate the flow map.

\subsection{Determination of $\beta$}   \label{SS:D_beta}


As discussed above, the choice of the sparsity constraint weight, $\beta$, in iterative SINDy plays a pivotal role in determining the algorithm's convergence and the quality of the obtained model. This parameter also decides the computational complexity.

A small value of $\beta$ may seem attractive as it can allow more terms to be included for better approximations. However, a small $\beta$ can lead to a large number of iterations (which results in a high computational cost) or even non-convergence, particularly when the dictionary rapidly expands to accommodate numerous terms. This can also result in a model that overfits the data, capturing noise or minor fluctuations in the system dynamics.

On the other hand, setting $\beta$ to a large value often leads to a zero-vector solution as discussed above, indicating that the algorithm essentially ignores the majority of candidate terms in the dictionary. This results in an overly simplistic model with low expressiveness and might not capture the complexity of the underlying dynamics. In this case, we may end up with a small number of iterations, equivalent to having a small dictionary, which can also lead to a poor approximation although the computation time is greately saved.

To strike the right balance, it is advisable to keep $\beta$ relatively small but not excessively so. This allows for some level of sparsity, preventing overfitting, while still capturing important terms that contribute to a more accurate approximation of the system's behavior. The exact choice of $\beta$ may vary depending on the specific dataset and application, and experimentation to find the optimal value is often necessary.


\section{Numerical Results} \label{S:Sim}

In this section, we present numerical results to evaluate the performance of iterative SINDy. We also compare iterative SINDy with conventional SINDy \cite{Brunton16}. To obtain sparse solutions, we employ Lasso, specifically using MATLAB's {\tt lasso} function for our simulations.

\subsection{Identification of Low-Dimensional Nonlinear Dynamics}

\begin{figure}[!t]
\begin{center} \hspace{-1.5em} 
\includegraphics[width=9cm]{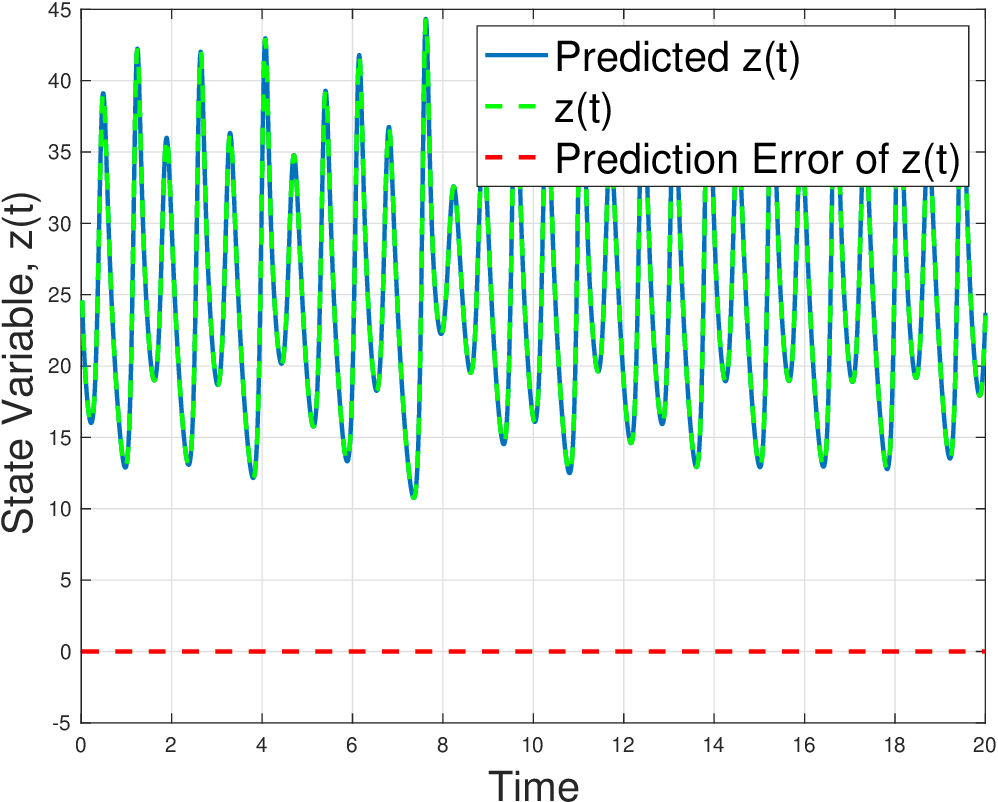} \\ 
(a) \\
\includegraphics[width=9cm]{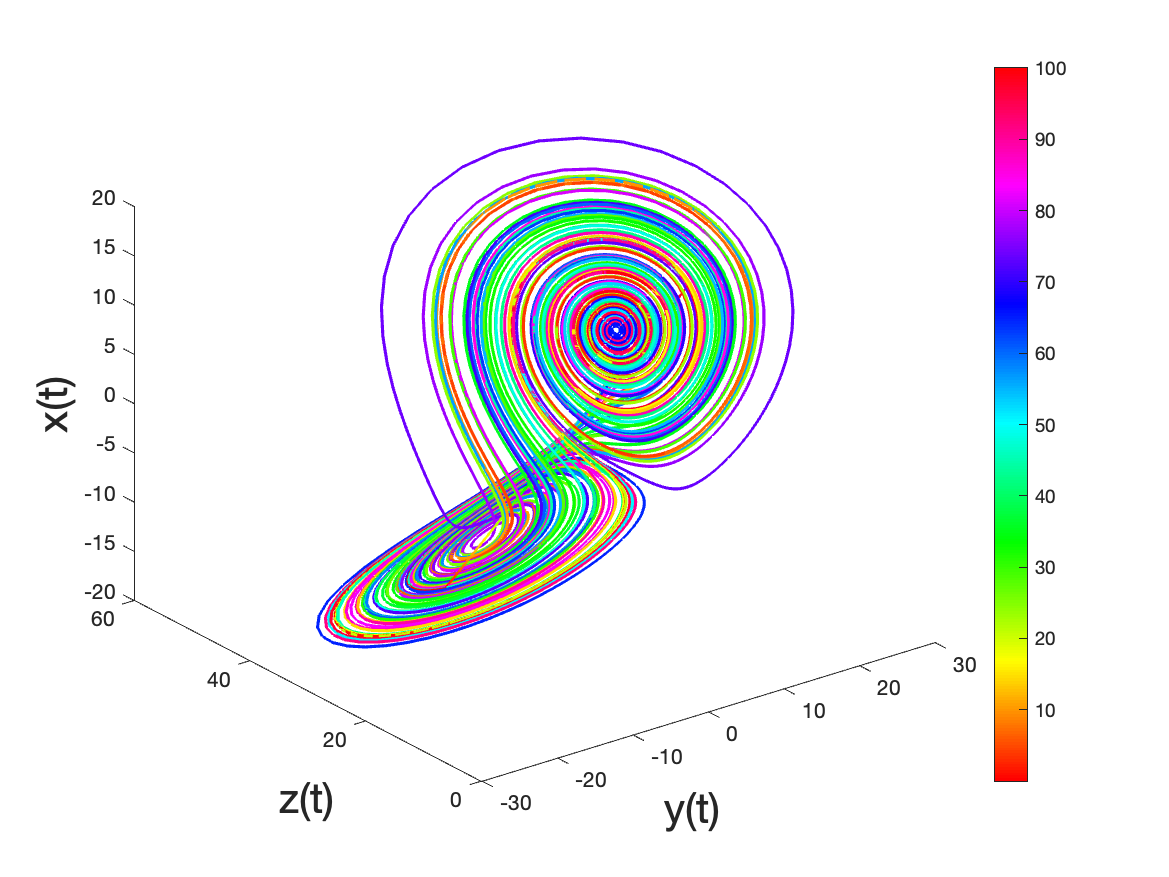} \\ 
(b) \\
\end{center}
\caption{Identification results using iterative SINDy: (a) predicted $z(t)$ and error signal; (b) a reproduced trajectory of 100 seconds.}
    \label{Fig:ISINDy}
\end{figure}

As shown in \cite{Brunton16}, SINDy can successfully model the Lorenz equation that is given by 
\begin{align}
    \dot{x} & = \sigma (y - x) \cr 
    \dot{y} & = x (\rho- z) - y \cr 
    \dot{z} & = x y - \alpha z,
        \label{EQ:Lorenz}
\end{align}
where $\sigma = 10$, $\rho = 28$, and $\alpha = 8/3$. Here, $\dot{x} = \frac{ {\rm d} x(t)}{{\rm d} t}$.
In addition, the initial values are $(x_0, y_0, z_0) = (-8, 7, 27)$. For simulations, the sampling time interval is set to ${\rm d} t = 0.01$ seconds. 
For iterative SINDy, the initial dictionary is set to $\cG_1 = \{1, x, y, z\}$ for all variables. After a few iterations, the final dictionary for each $x_n$ becomes $\cG_S = \{x,y,xy\}$ for $x$, $\cG_S = \{y,xy,x,xz\}$ for $y$, and $\cG_S = \{z,xy,y^2,x^2\}$ for $z$. The coefficients corresponding to the terms that do not appear in the original dynamics in \eqref{EQ:Lorenz} are negligible. For example, the coefficients of $x^2$ and $y^2$ in $\mathcal{G}_S$ for $z$ are less than $10^{-4}$. 
In Fig.~\ref{Fig:ISINDy}, it is shown that the nonlinear dynamics of the Lorenz equation can be accurately identified using iterative SINDy. Specifically, the predicted state variable $z(t)$ is nearly identical to the actual $z(t)$, as shown by the error signal in Fig.~\ref{Fig:ISINDy} (a), which is almost zero. Furthermore, the trajectory of the Lorenz equation is reproduced, as depicted in Fig.~\ref{Fig:ISINDy} (b), successfully capturing the attractor dynamics, similar to SINDy in \cite{Brunton16}.

\begin{figure}[!t]
\begin{center} \hspace{-1.5em} 
\includegraphics[width=9cm]{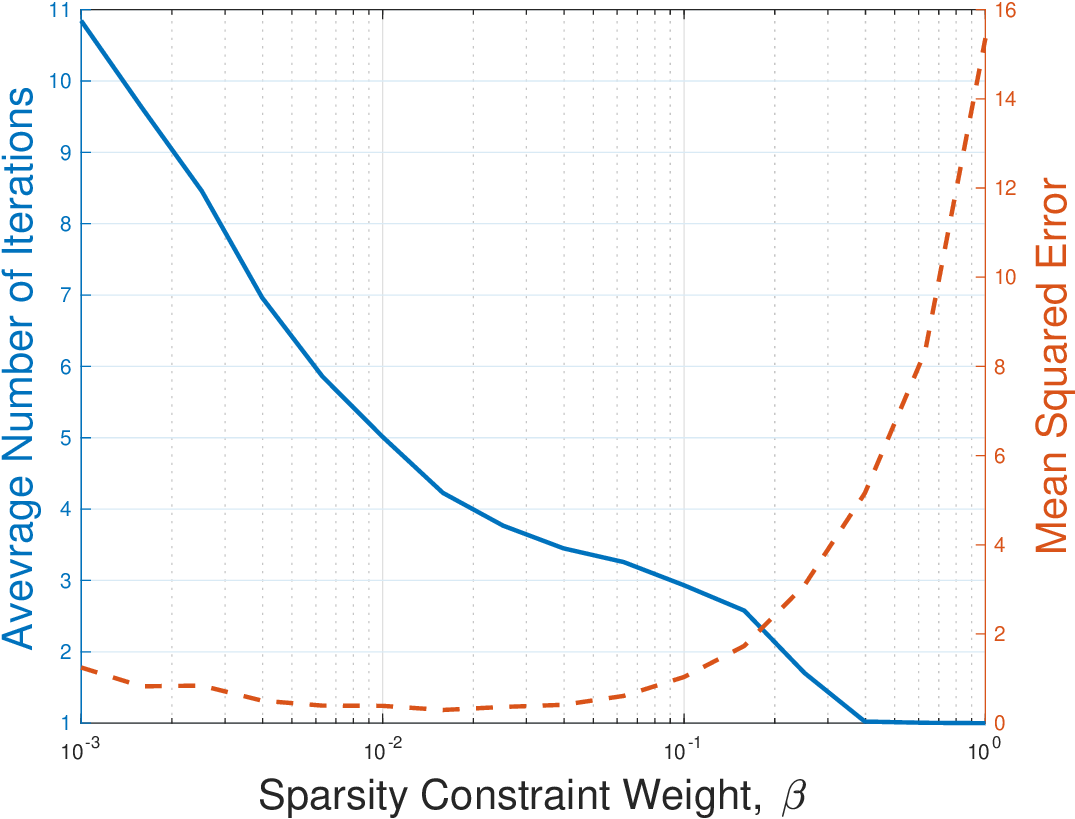}
\end{center}
\caption{Average number of iterations and MSE as functions of  $\beta$.}
    \label{Fig:converg}
\end{figure}

In order to see the convergence behaviors of iterative SINDy, we consider the following signal for approximation:
\be
w(t) = s(t+ {\rm d} t)  + u (t),
\ee 
where $s(t) = x(t) + y(t) + z(t)$ is the sum signal of all the state variables in \eqref{EQ:Lorenz} and $u(t) \sim \cN (0, \sigma^2)$ is the noise. Clearly, $w(t)$ is a noisy version of the sum of the future states, which can be approximated by a linear combination of candidate functions of the current states, $x(t)$, $y(t)$, and $z(t)$, using iterative SINDy.
For simulation, we assume that the variance of the noise, $\sigma^2$, is determined as follows:
\be 
\sigma^2 = \frac{\frac{1}{T} \int_0^T |s(t)|^2 {\rm d} t}{\rm SNR},
\ee 
where ${\rm SNR}$ represents the signal-to-noise ratio (SNR), which is set to 20 dB. In Fig.~\ref{Fig:converg}, the mean of the number of iterations is shown for different values of $\beta$ with $S = 20$. We also show the mean-squared error (MSE), which is $\uE[|w(t) - \hat w(t)|^2]$, where $\hat w(t)$ is the linear combination of the candidate functions that are selected by iterative SINDy. This mean  is computed over 400 runs.
It is shown that the number of iterations increases as $\beta$ decreases, which is expected from Section~\ref{S:Analysis}. In general, as $\beta$ decreases, the sparsity constraint becomes weaker, and more terms can be used for a better approximation. However, as shown in Fig.~\ref{Fig:converg}, having more terms does not necessarily lead to a lower MSE. 
It implies that while iterative SINDy can include more terms in the model under weaker sparsity constraints, not all of these additional terms may contribute positively to reducing the prediction error. 
Some of these terms might capture noise or lead to overfitting  as discussed in Subsection~\ref{SS:D_beta}, which can increase the MSE. Consequently, it is crucial to carefully select an appropriate value of $\beta$ that keeps the number of iterations small while ensuring a good approximation.

\subsection{Application to Multivariate Time-Series Data}

\begin{figure}[!t]
\begin{center} \hspace{-1.5em} 
\includegraphics[width=9cm]{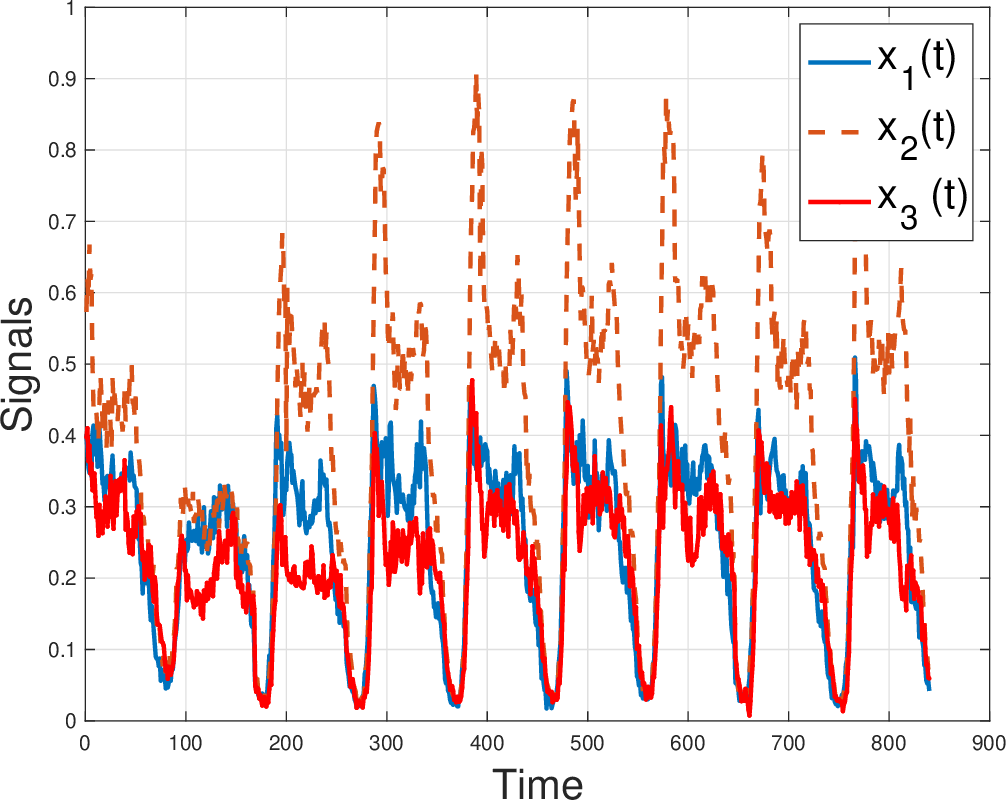}
\end{center}
\caption{Traffic volume data from \cite{Zhao19} as multivariate time-series.}
    \label{Fig:Traffic}
\end{figure}

With the known dynamics of the Lorenz equation, we have readily verified the effectiveness of iterative SINDy as an alternative to SINDy. In particular, as shown in \eqref{EQ:Lorenz}, a few monomial terms are sufficient to model the dynamics in both SINDy and iterative SINDy. However, in practice, when a multivariate time-series dataset is provided for modeling using a nonlinear dynamical system, the dynamics are often more complex, and the underlying equations are typically unknown. To explore the capabilities of iterative SINDy in real-world applications, we have utilized traffic volume data from \cite{Zhao19}. This dataset consists of measurements taken at 36 locations, with each sequence containing 840 samples. Fig.~\ref{Fig:Traffic} presents the traffic volume data from three different locations (out of the 36 available locations) that constitute the state vector. A nonlinear dynamical system is assumed to model the traffic volumes as its state.

\begin{figure}[!t]
\begin{center} \hspace{-1.5em} 
\subfigure[Predicted 3rd signal by SINDy]{\label{figS1_ax}\includegraphics[width=0.45\textwidth]{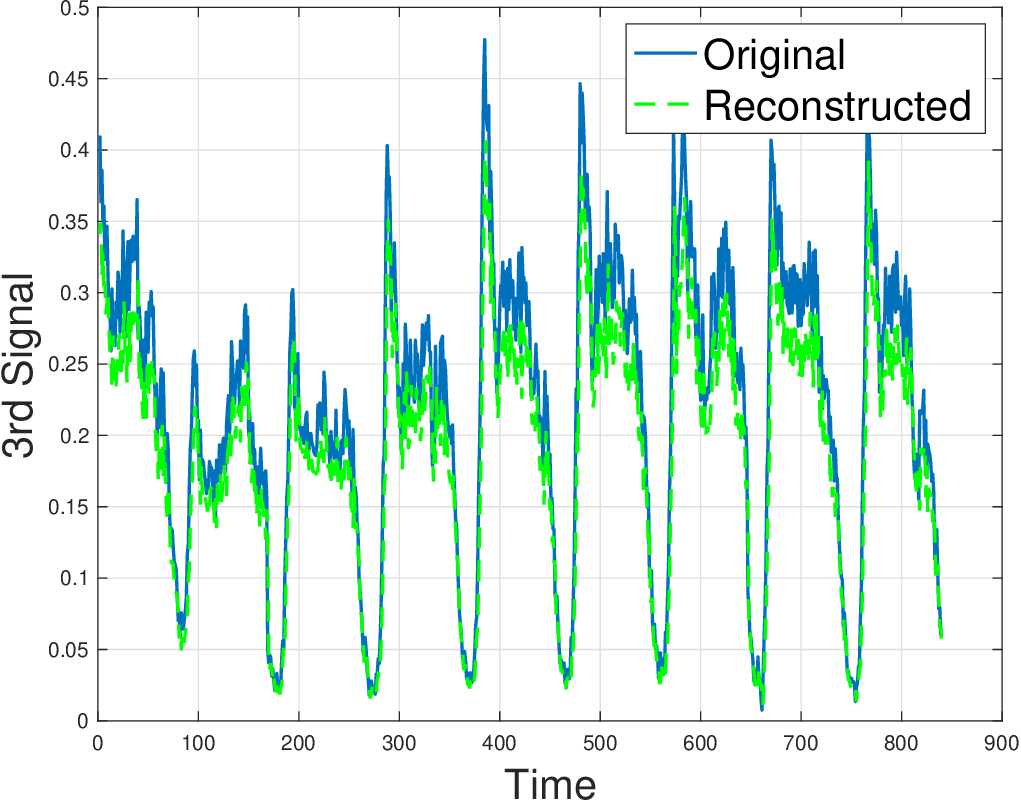}}
\subfigure[Predicted 3rd signal by iterative SINDy]{\label{figS1_ax}\includegraphics[width=0.45\textwidth]{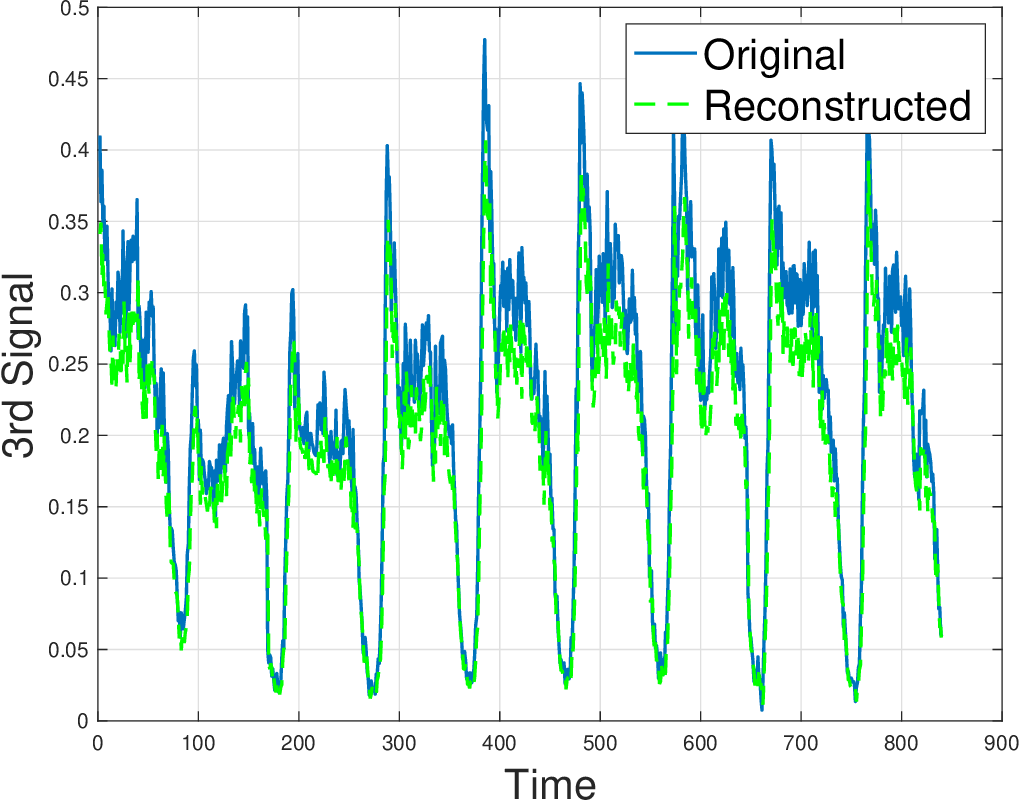}} \\
\subfigure[Error]{\label{figS1_bx}\includegraphics[width=0.24\textwidth]{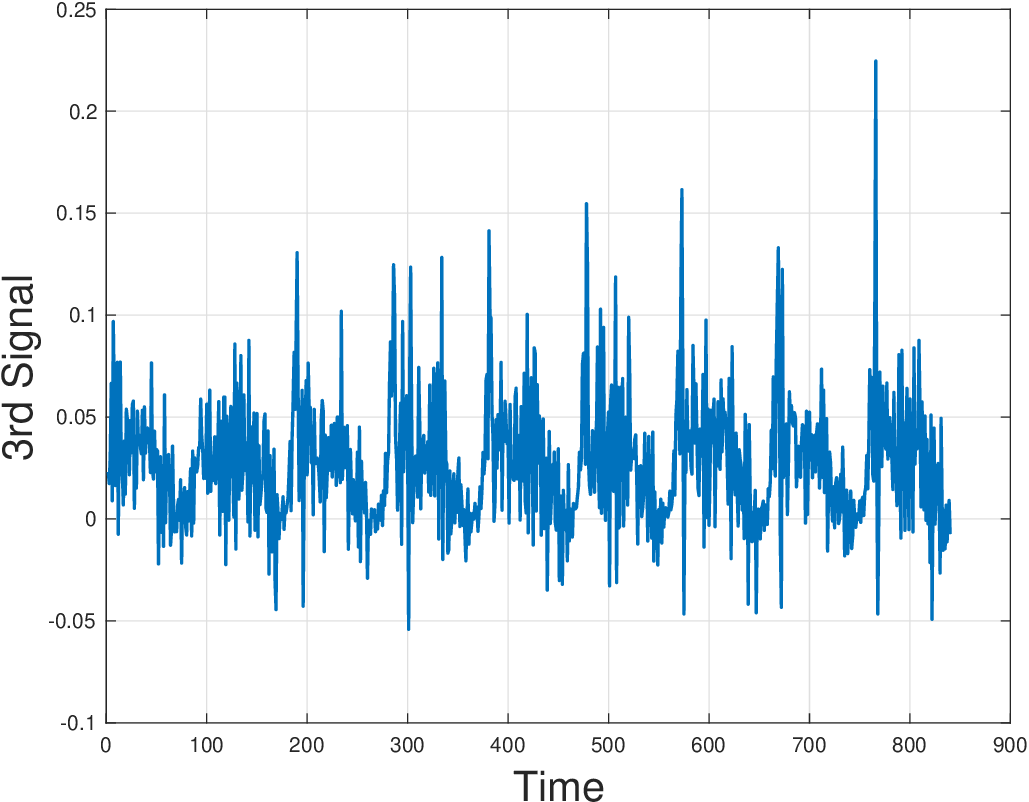}} 
\subfigure[Error]{\label{figS1_bx}\includegraphics[width=0.24\textwidth]{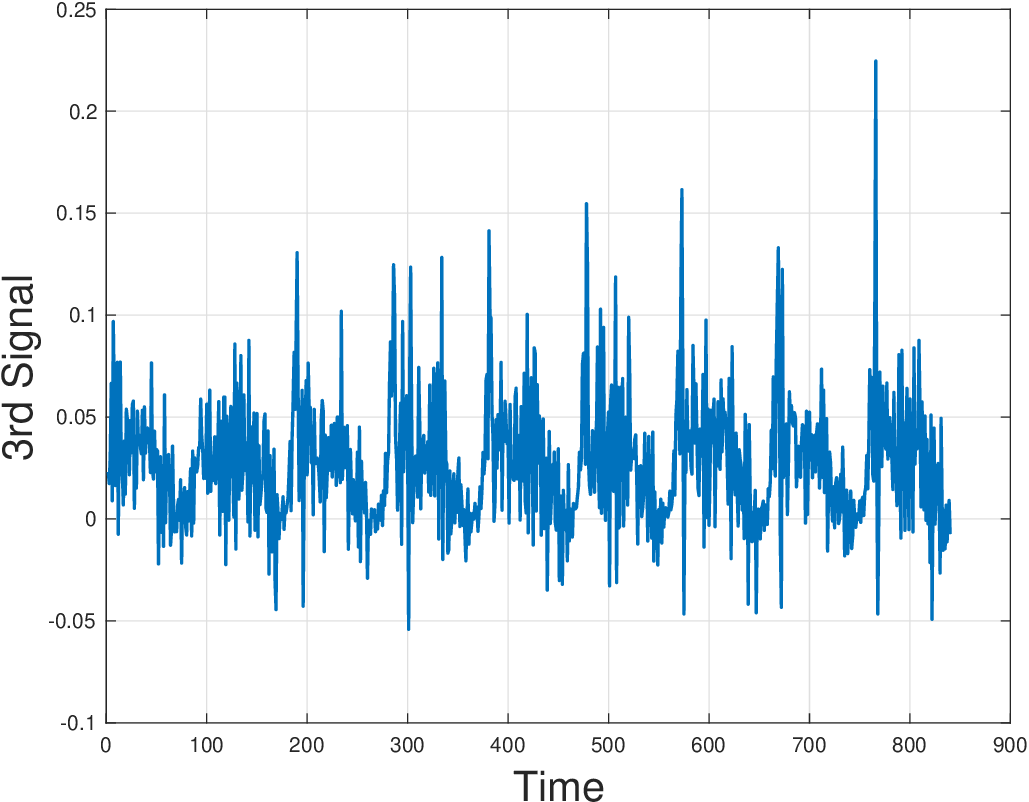}} 
\end{center}
\caption{Modeling by the SINDy and iterative SINDy approaches when $N = 3$, $S = 4$, and $\beta = 0.01$: (a) The 3rd data and its prediction by SINDy; (b) The 3rd data and its prediction by iterative SINDy; (c) Error signal of SINDy; (d) Error signal of iterative SINDy.}
    \label{Fig:SINDy}
\end{figure}

We apply the SINDy and iterative SINDy approaches to the dataset in Fig.~\ref{Fig:Traffic}, where $N = 3$ sensor locations out of 36 are shown. In the related simulations hereafter, the number of input variables, $N$, is now the number of selected sensor locations. Thus, $N \le 36$. 
In SINDy, it is assumed that the dictionary is obtained by expanding the unity set,  $\cG_1 = \{1, x_1, \ldots, x_N \}$, multiple times (say $S$ times to compare it with iterative SINDy, where $S$ is the maximum number of iterations). Hence, to represent the size of the dictionary for SINDy in this case, we use $K(N, S)$ instead of just $K$.
With $S = 4$, the size of dictionary for SINDy is shown for different numbers of $N$ in Table~\ref{TBL:SD_K}, which demonstrates that the computational complexity might be prohibitively high if the input dimension, $N$, is large \cite{Baddoo22}. 
For example, when $N = 4$, the size of the resulting dictionary becomes $K(N,S) = 126$. 
On the other hand, the dictionary is expanded and compressed through iterations in iterative SINDy with a maximum number of iterations of $S$ to keep a reasonable size of dictionary. 
While both SIND and iterative SINDy can have different computational complexity, it is shown that  both the approaches provide near identical results with the traffic data in \cite{Zhao19} as illustrated in Fig.~\ref{Fig:SINDy}.

\begin{table}[!t]
\caption{The size of dictionary for SINDy obtained by expanding the unity set  $S = 4$ times for different values of $N$.}
\begin{tabular}{|c||c|c|c|c|c|c|c|c|c|}
\hline
$N$ & 2 & 4 & 6 & 8 & 10 & 12 & 14 & 16 \\
\hline
$K$ & 21 & 126 & 462 & 1287 & 3003 & 6188 & 11628 & 20349 \\
\hline
\end{tabular}
 \label{TBL:SD_K}
\end{table}

For further comparison, we consider the modeling error, which is defined as
\be
{\rm Error} = \frac{1}{T} \sum_{t=0}^{T-1} \sum_{n=1}^N |x_n (t+1) - \bpsi (\bx(t))^\rT \bb_n|^2 ,
    \label{EQ:merror}
\ee 
where $\bpsi(\cdot)$ and $\bb_n$ are obtained by solving Lasso in SINDy or iterative SINDy. Additionally, we consider the total number of monomial terms used in $\bpsi(\cdot)$, which determines the model's order or complexity.

\begin{figure}[!h]
\begin{center} \hspace{-1.5em} 
\subfigure[Error vs $\beta$]{\label{figS1_ax}\includegraphics[width=0.40\textwidth]{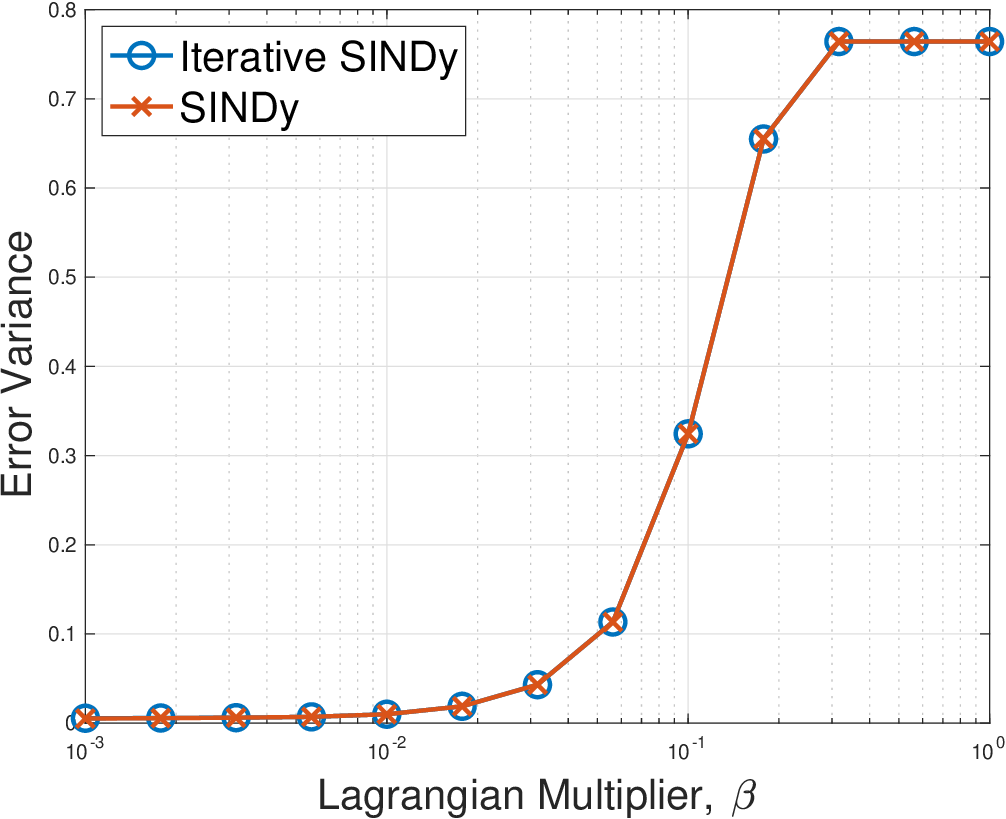}}
\subfigure[Computation time vs $\beta$]{\label{figS1_bx}\includegraphics[width=0.40\textwidth]{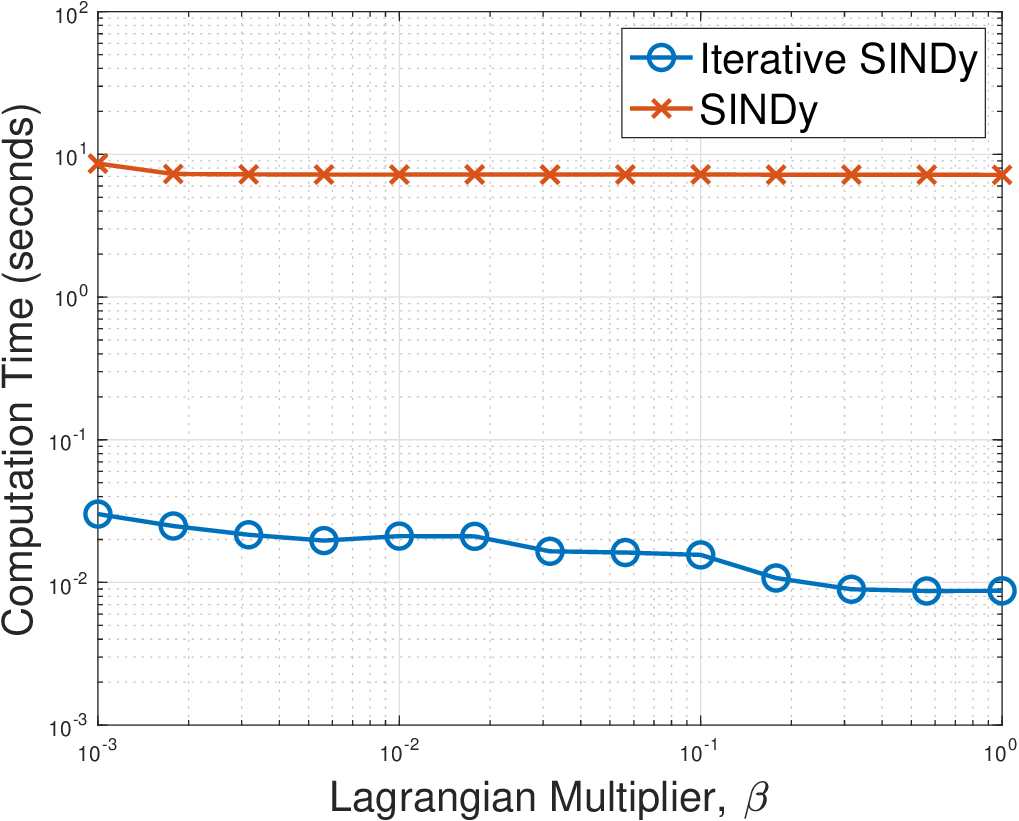}}  \\
\subfigure[Sum of orders vs $\beta$]{\label{figS1_bx}\includegraphics[width=0.40\textwidth]{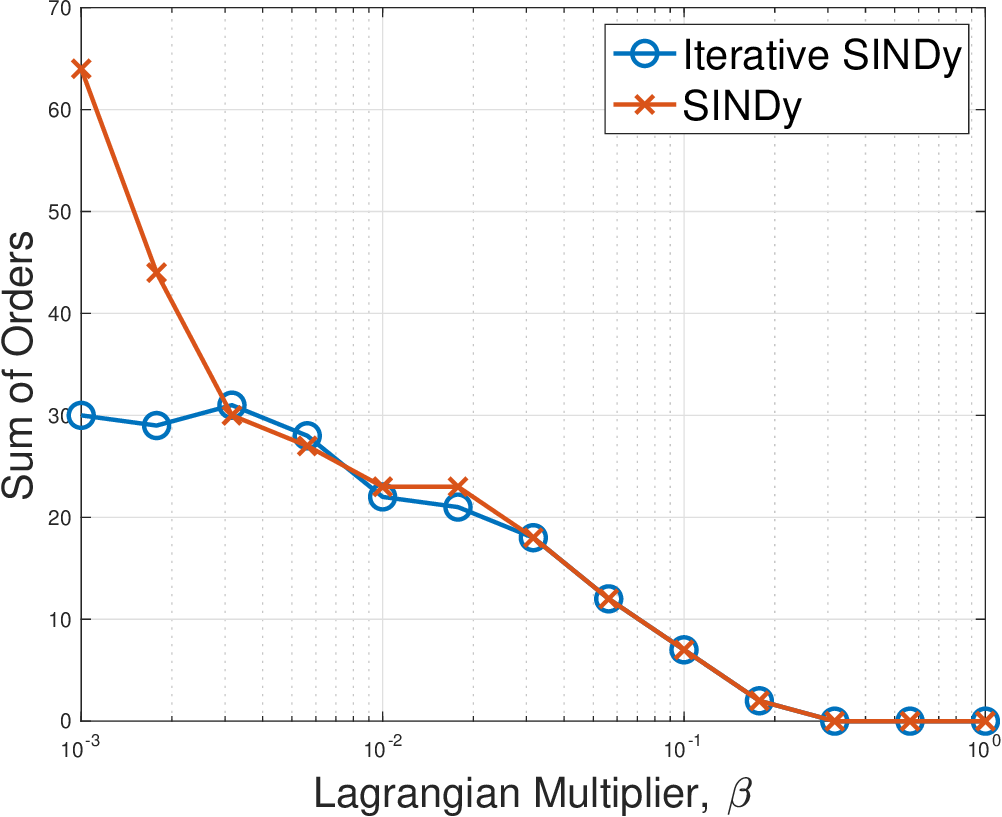}} 
\end{center}
\caption{Performance of SINDy and iterative SINDy for varying sparsity constraint weight $\beta$ when $N = 6$ and $S = 4$: (a) Approximation error variance as a function of $\beta$; (b) Computation time as a function of $\beta$; (c) Sum of model orders as a function of $\beta$.}
    \label{Fig:plt_traf2}
\end{figure}

To compare the performance difference between SINDy and iterative SINDy, we consider various values of the sparsity constraint weight, i.e., $\beta$. 
The results are shown in Fig.~\ref{Fig:plt_traf2} for the case of $N = 6$ and $S = 4$. Both approaches provide nearly identical performance in terms of modeling error, as displayed in Fig.~\ref{Fig:plt_traf2} (a). 
We also note that $\beta$ needs to be sufficiently small to achieve a low modeling error. For instance, a value of $\beta \le 0.01$ is necessary for a low modeling error.
While both approaches yield similar modeling errors, they exhibit different computation time results, which were measured using MATLAB's {\tt tic} and {\tt toc} commands. In SINDy, the computation time primarily depends on the Lasso solver and tends to slightly decrease with increasing $\beta$, as shown in Fig.~\ref{Fig:plt_traf2} (b). On the other hand, in iterative SINDy, the dictionary size is constrained during each iteration, leading to a more substantial reduction in computation time with increasing $\beta$. This illustrates that the iterative SINDy approach can be more computationally efficient than the conventional SINDy approach.
It is noteworthy that the total number of monomial terms (or orders) for $N = 6$ state variables decreases with increasing $\beta$, as shown in Fig.~\ref{Fig:plt_traf2} (c). This indicates that a more compact model can be obtained with larger values of $\beta$. Importantly, the modeling error increases with $\beta$, as fewer monomial terms are used, as illustrated in Fig.~\ref{Fig:plt_traf2} (a) and (c). However, iterative SINDy can provide a more compact model than SINDy when $\beta$ is small (which is necessary for achieving a reasonably small modeling error) without degrading the modeling quality.

\begin{figure}[!h]
\begin{center} \hspace{-1.5em} 
\subfigure[Error vs $S$]{\label{figS1_ax}\includegraphics[width=0.4\textwidth]{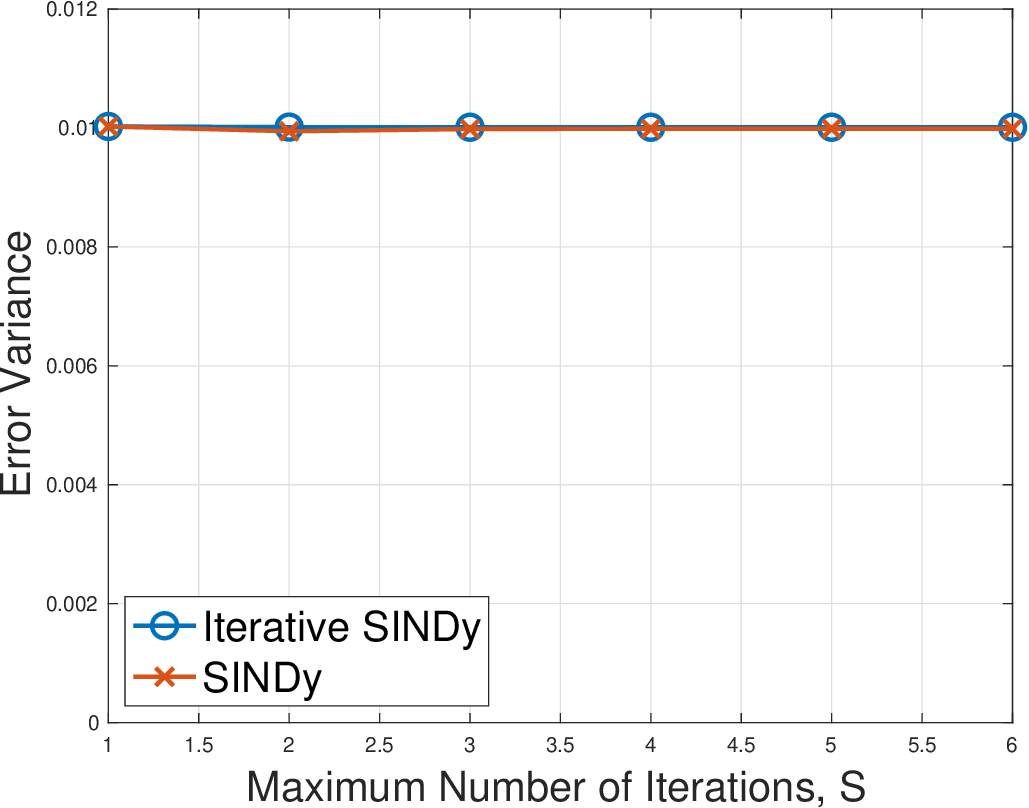}}
\subfigure[Computation time vs $S$]{\label{figS1_bx}\includegraphics[width=0.4\textwidth]{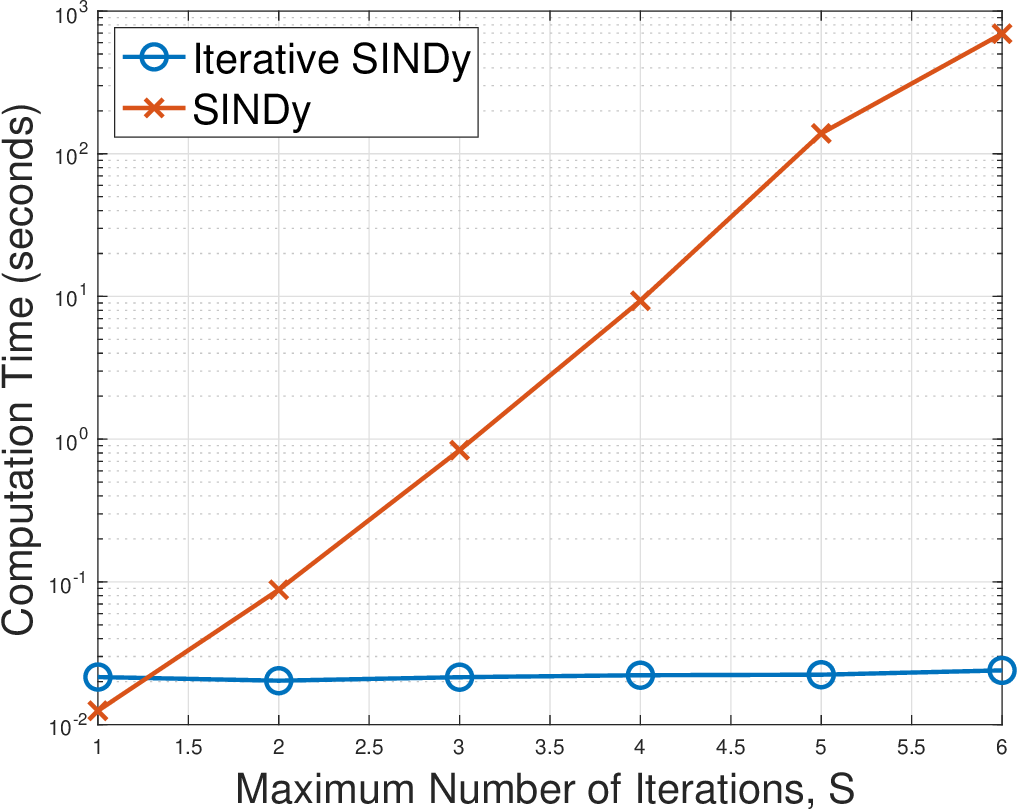}}  \\
\subfigure[Sum of orders vs $S$]{\label{figS1_bx}\includegraphics[width=0.4\textwidth]{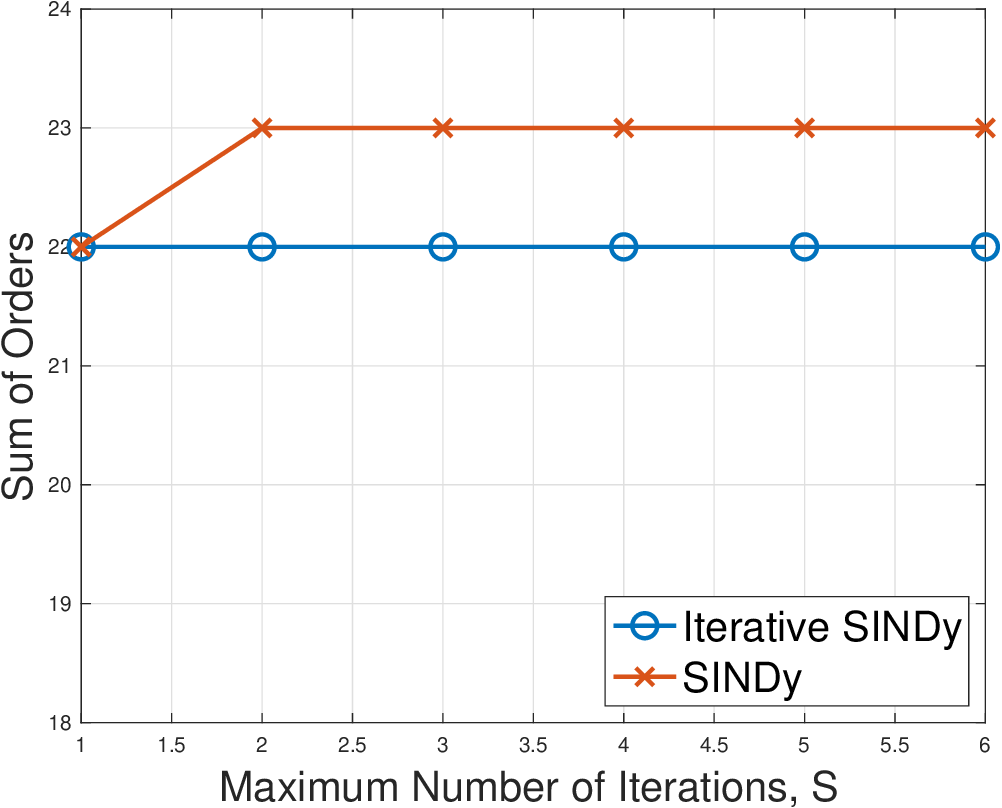}} 
\end{center}
\caption{Performance of SINDy and Iterative SINDy for varying $S$ when $N = 6$  and $\beta = 0.01$: (a) Approximation error variance as a function of $S$; (b) Computation time as a function of $S$; (c) Sum of model orders as a function of $S$.}
    \label{Fig:plt_traf1}
\end{figure}

The impact of the maximum number of iterations, $S$, on the performance is depicted in Fig.~\ref{Fig:plt_traf1} when $N = 6$  and $\beta = 0.01$. Note that the size of the dictionary of conventional SINDy, $K (N,S)$, increases with $S$ as a function of $S$.
As shown in Fig.~\ref{Fig:plt_traf1} (a), both the approaches have almost identical modeling error, which tends to be independent of $S$, because a sufficient model order for reasonable approximation can be found with a small $S$ as illustrated in Fig.~\ref{Fig:plt_traf1} (c).  
Note that since  the size of the dictionary exponentially increases with $S$ in SINDy, the computation time significantly increases with $S$. On the other hand, the size of the dictionary in iterative SINDy does not necessarily increase with $S$, because the iteration can stop when the stopping criterion is met. Thus, as shown in Fig.~\ref{Fig:plt_traf1} (b), iterative SINDy can have a much lower computation time than SINDy as $S$ increases.

Although a large dictionary results in  high computational complexity in SINDy, SINDy typically relies on a large dictionary due to the lack of prior knowledge about which candidate functions are effective. In other words, in SINDy, the use of a large dictionary is primarily based on the hope that it may include some useful candidate functions, making the computational cost unavoidable.
On the other hand, the iterative SINDy approach can find a dictionary by expanding and compressing through iterations, meaning that  the dictionary size can be adjusted for a specific dataset. Furthermore, the computation time of iterative SINDy can be almost independent of $S$ as shown in Fig.~\ref{Fig:plt_traf1} (b), indicating that once a suitable model is selected, no further iterations are necessary.

\begin{figure}[!h]
\begin{center} \hspace{-1.5em} 
\subfigure[Error vs $N$]{\label{figS1_ax}\includegraphics[width=0.4\textwidth]{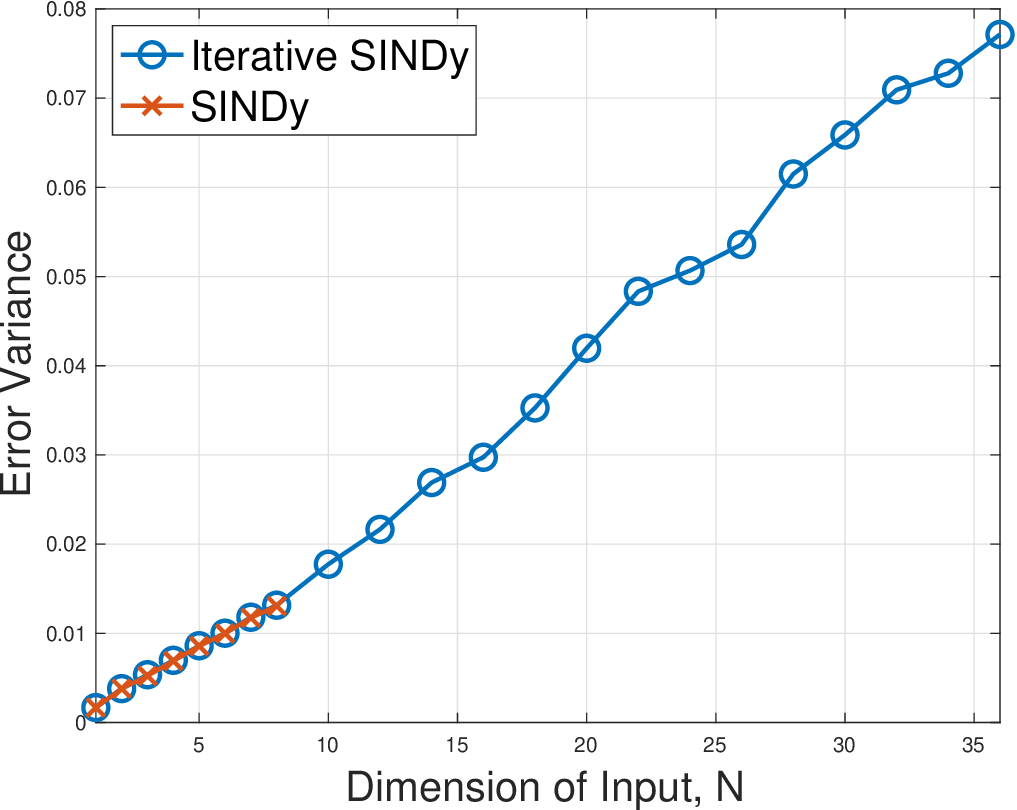}}
\subfigure[Computation time vs $N$]{\label{figS1_bx}\includegraphics[width=0.4\textwidth]{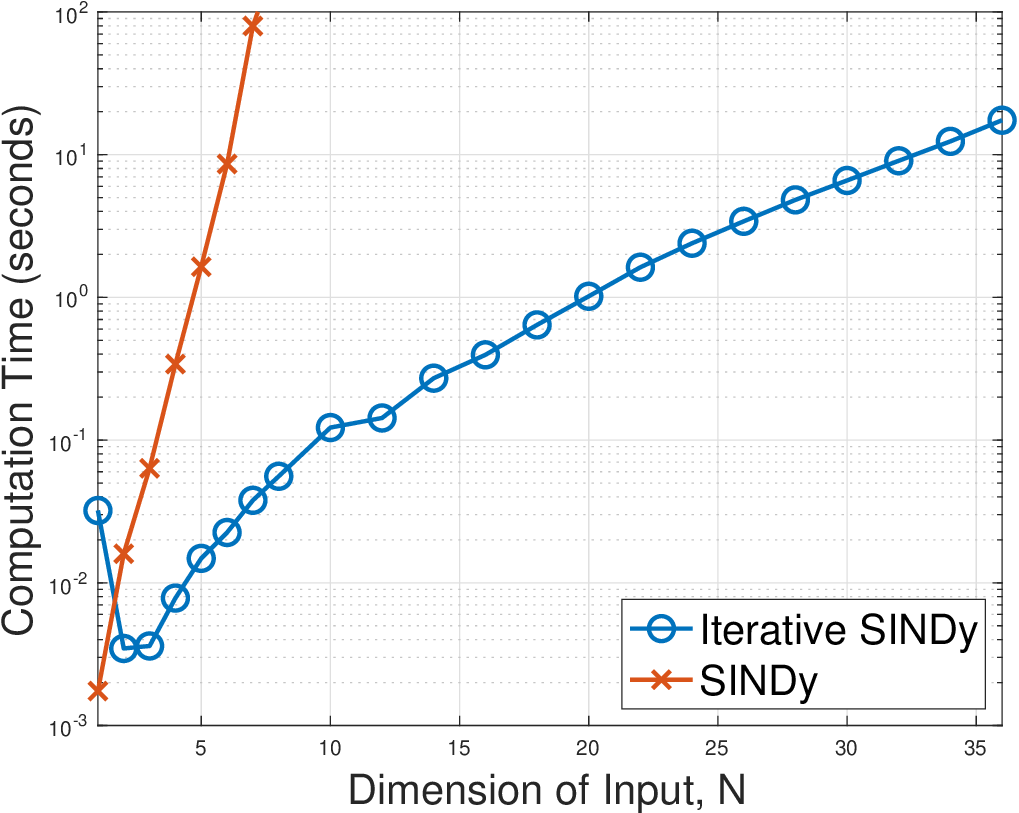}}  \\
\subfigure[Sum of orders vs $N$]{\label{figS1_bx}\includegraphics[width=0.4\textwidth]{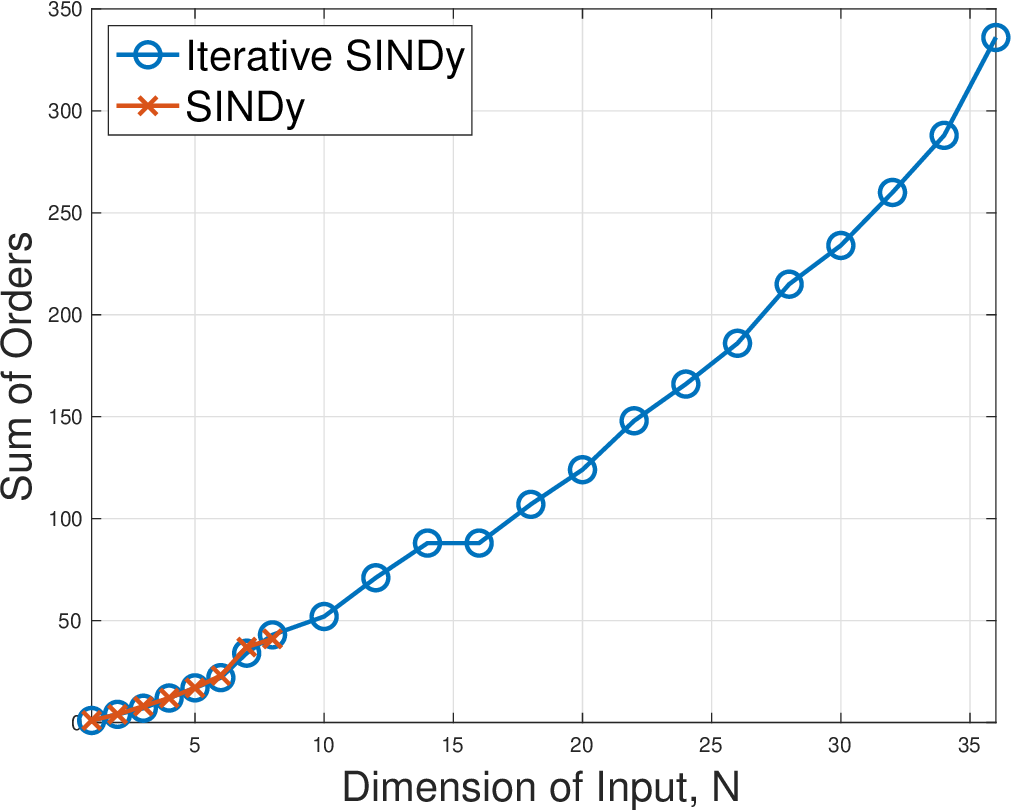}} 
\end{center}
\caption{Performance of SINDy and Iterative SINDy for varying input size $N$ when $S = 4$ and $\beta = 0.01$: (a) Approximation error variance as a function of $N$; (b) Computation time as a function of $N$; (c) Sum of model orders as a function of $N$.}
    \label{Fig:plt_traf3}
\end{figure}

SINDy faces challenges when dealing with high-dimensional input data (large $N$) \cite{Baddoo22} due to the resulting increase in dictionary size, which leads to higher computational costs. In contrast, iterative SINDy dynamically adjusts the dictionary size through expansion and compression, making it particularly suitable for modeling high-dimensional data. To illustrate this, we analyze the impact of increasing $N$ on the performance, as shown in Fig.~\ref{Fig:plt_traf3} for $S = 4$ and $\beta = 0.01$. As depicted in Fig.~\ref{Fig:plt_traf3} (a) and (c)\footnote{Note that the results of SINDy are obtained only for small values of $N$, i.e., $N \in \{1, \ldots, 8\}$ due to prohibitively high computational complexity, while large values of $N$ up to 36 are used for iterative SINDy.}, the performance of iterative SINDy remains competitive with that of SINDy across a wide range of $N$, but the computational complexity of iterative SINDy is significantly lower than that of SINDy as $N$ increases. 
As mentioned earlier, this highlights that iterative SINDy is particularly well-suited for handling high-dimensional datasets.

It is noteworthy that both the modeling error and the number of monomial terms increase with $N$, as depicted in Fig.~\ref{Fig:plt_traf3} (a) and (c). While the linear increase in modeling error is expected (due to the utilization of more terms, as indicated in \eqref{EQ:merror}), it is interesting that the growth in the number of monomial terms is slightly faster than linear. This observation suggests a relationship between the state variables, with more cross terms being included as $N$ increases.

\section{Conclusions}   \label{S:Conc}

In this paper, we introduced iterative SINDy, a novel approach for identifying governing equations from time-series datasets. By iteratively expanding and compressing the dictionary of candidate functions, we have demonstrated that iterative SINDy can adapt to a given time-series dataset and accurately approximates the underlying dynamics, like traditional SINDy. Our analysis of convergence properties reaffirms the robustness of this approach. In addition, through a series of numerical experiments, we showcased iterative SINDy's performance compared to SINDy. The results validate its ability to maintain comparable approximation accuracy while significantly reducing computational complexity, particularly in scenarios involving high-dimensional input data.
However, it is worth noting that while our study focuses on the dimension of candidate functions in terms of computational complexity, the computational bottleneck may also arise from the growing number of time points, $T$, even if the number of bases is adequate. Addressing this aspect could be an important direction for future research. Furthermore, considering the extension to continuous systems involving partial differential equations in the future would be valuable.

Iterative SINDy offers a practical solution to the challenges of model identification and complexity, rendering it a valuable tool across various applications. Its adaptability and capacity to handle high-dimensional data make it a promising approach. However, the proposed approach for iterative SINDy also has various limitations. For instance, by limiting our dictionary to monomials, we may only approximate the model when dealing with harmonic components (such as in the case of a double pendulum). While it is feasible to include sine and cosine terms in the dictionary, as demonstrated in \cite{Brunton16}, this may compromise the closure property of the dictionary in terms of monomials, and the current convergence analysis may not hold. In the future, we hope to explore alternative approaches that can accommodate harmonic components while providing convergence guarantees.


\bibliographystyle{ieeetr}
\bibliography{Koopman}

\end{document}